\RequirePackage{fix-cm}
\documentclass[english,notitlepage,prl,reprint,unsortedaddress,longbibliography]{revtex4-1}
\usepackage[latin9]{inputenc}
\usepackage{textcomp}
\usepackage{amsmath}
\usepackage{amssymb}
\usepackage{graphicx}

\makeatletter

\DeclareFontEncoding{LGR}{}{}
\DeclareRobustCommand{\greektext}{%
  \fontencoding{LGR}\selectfont\def\encodingdefault{LGR}}
\DeclareRobustCommand{\textgreek}[1]{\leavevmode{\greektext #1}}
\ProvideTextCommand{\~}{LGR}[1]{\char126#1}

\makeatother

\usepackage{babel}
\usepackage{listings}

\begin{document}

\title{Spin blockades to relaxation of hot multi-excitons in nanocrystals }

\author{Tufan Ghosh}

\affiliation{Fritz Haber Center for Molecular Dynamics and the Institute of Chemistry,
The Hebrew University of Jerusalem, Jerusalem 9190401, Israel.}

\author{Marcel D. Fabian}

\affiliation{Fritz Haber Center for Molecular Dynamics and the Institute of Chemistry,
The Hebrew University of Jerusalem, Jerusalem 9190401, Israel.}

\author{Joanna Dehnel}

\affiliation{Schulich Faculty of Chemistry, Solid State Institute, Russell Berrie
Nanotechnology Institute, Nancy and Stephen Grand Technion Energy
Program, and Department of Materials Science and Engineering, Technion
Israel Institute of Technology, Haifa 3200003, Israel.}

\author{Efrat Lifshitz}

\affiliation{Schulich Faculty of Chemistry, Solid State Institute, Russell Berrie
Nanotechnology Institute, Nancy and Stephen Grand Technion Energy
Program, and Department of Materials Science and Engineering, Technion
Israel Institute of Technology, Haifa 3200003, Israel.}
\email{efrat333@gmail.com}

\author{Roi Baer}

\affiliation{Fritz Haber Center for Molecular Dynamics and the Institute of Chemistry,
The Hebrew University of Jerusalem, Jerusalem 9190401, Israel.}
\email{roi.baer@huji.ac.il}

\author{Sanford Ruhman}

\affiliation{Fritz Haber Center for Molecular Dynamics and the Institute of Chemistry,
The Hebrew University of Jerusalem, Jerusalem 9190401, Israel.}
\email{sandy@mail.huji.ac.il}

\begin{abstract}
The rates of elementary photophysical processes in nanocrystals, such
as carrier cooling, multiexciton generation, Auger recombination etc.,
are determined by monitoring the transient occupation of the lowest
exciton band. The underlying assumption is that hot carriers relax
rapidly to their lowest quantum level. Using femtosecond transient
absorption spectroscopy in CdSe/CdS nanodots we challenge this assumption.
Results show, that in nanodots containing a preexisting cold ``spectator
exciton'', \emph{only half of the photoexcited electrons }relax directly
to the band-edge and the rest are blocked in an excited state due
to Pauli exclusion. Full relaxation occurs only after \textasciitilde 15
ps, as the blocked electrons flip spin. This requires review of numerous
studies unaware of this ubiquitous and novel effect, which may facilitate
hot carrier energy utilization as well.
\end{abstract}
\maketitle

\section{Introduction}

Understanding how excess photon energy is dissipated after nanocrystal
(NC) photoexcitation is essential for utilizing this energy in nanodot
based solar cells or photo-detectors \citep{Koenig2010,Pandey2009,Saeed2014,Konstantatos2013,Williams2013,li2017slow}.
Decades of ultrafast investigation show that inter-band photoexcitation
of quantum dots is followed by rapid relaxation of hot carriers to
the quantized band edge (BE) states within one or two picoseconds\citep{woggon1996ultrafast,Nozik2001,Guyot-Sionnest1999,Klimov1999,Klimov2000,Sewall2006,Kambhampati2011,Cooney2007a,Gdor2012,gdor2013novel,Spoor2015,Spoor2017}.
Due to the large oscillator strength and low degeneracy of the BE
exciton transition, evolution in its intensity and spectrum have played
a pivotal part in probing quantum dot exciton cooling. These allegedly
start with bi-exciton spectral shifts while carriers are hot, changing
to a photoinduced bleach (PIB) due to state filling once the exciton
relaxes. Accordingly, kinetics of this PIB buildup has served to characterize
the final stages of carrier cooling \citep{Klimov1999}, and its amplitude
per cold exciton provides a measure of underlying state degeneracy
\citep{Efros1982,Esch1990}. 

Hot multi-excitons (MX), generated for instance through sequential
multi-photon absorption of femtosecond pulses, \citep{Klimov2007,Kambhampati2012},
add a new relaxation process to this scenario \citep{Klimov2000b}.
Auger recombination (AR) reduces an N exciton state to N-1 plus heat,
initially deposited in the remaining carriers and later transferred
to the lattice. Depending on N, this shortens the lifetime of MXs
relative to a single exciton by two to three orders of magnitude.
Again, investigation of AR dynamics is based on the amplitude and
decay kinetics of the BE bleach with interpretation based on the following
assumptions: 1) that ultrafast cooling of hot excitons leads directly
to occupation of the lowest electron and hole states (in accordance
with the lattice temperature and the state degeneracy), and 2) that
aside from mild spectral shifts induced in the remaining band edge
transitions, after carrier cooling is over the BE PIB increases linearly
with N until the BE states are full (again dictated by state degeneracy). 

To test these assumptions, three pulse saturate-pump-probe experiments
were conducted in our lab, measuring fs transient absorption (TA)
of PbSe NCs in the presence and absence of single cold spectator excitons
\citep{Gdor2015}. This method hinges on the separation of timescales
between AR and radiative recombination, the latter being much slower.
Given the vast absorption cross sections of NCs, \citep{Leatherdale2002,Osborne2004,Jasieniak2009}
it is easy to excite nearly all particles in a sample with at least
one photon even with ultrashort pulses. The rapid ensuing AR leads
to a uniform population of NCs, all populated with a single cold exciton,
within a fraction of a nanosecond. One can then probe the effect of
a second exciton by comparing equivalent ultrafast pump-probe experiments
in the samples with or without spectator excitons. Surprisingly, the
BE bleach induced by a single relaxed exciton was found to be significantly
larger than that induced by a second exciton which was added by above
BE photoexcitation and left to cool down for 1-2 ps \citep{Gdor2015}.
This finding was also shown to be consistent with the fluence dependent
BE PIB saturation when exciting well above the optical band gap (BG)
via comparison with simulations.

To unveil the cause of this discrepancy, the same approach is applied
here to MXs in CdSe nanocrystals. This serves to test the generality
of the results in lead salt. Due to the reduced band edge electron
state degeneracy, each exciton has a much larger effect on the BE
PIB in CdSe NCs, which simplifies analysis and assignment. Numerous
investigations of CdSe NC photophysics have established: 1) that the
BE exciton absorption is practically insensitive to hole state occupancy,
allegedly due to the high density of valence bands in this material,\citep{Klimov1999,Sewall2006}
2) that the BE transition in CdSe NCs reflects only a two-fold spin
related degeneracy in the electron states \citep{Efros2000}. Accordingly
one relaxed exciton should bleach one-half of the initial BE absorption
band, and the second erase it entirely. 3) that the much higher effective
heavy hole mass in CdSe and involvement in spin-orbit coupling leads
to much faster cooling of holes (\textless psec) relative to electrons
\citep{Klimov1999}. Furthermore, the rapid electron relaxation measured
in CdSe NCs is assigned to Auger cooling where excess electron energy
transfers to the hole, and is then degraded to phonons \citep{Klimov1999,Hendry2006,Cooney2007a}.

Contrary to the assumptions outlined above, our results show that
adding a hot exciton to a relaxed singly occupied CdSe NC bleaches
only half of the remaining BE absorption once the initial carrier
cooling is over. It is further demonstrated that incomplete bleaching
by the second exciton is due to hitherto unrecognized random spin
orientation conflicts between the two sequentially excited conduction
electrons in part of the crystallites. The presence of this effect
both in lead salts and in CdSe NCs demonstrates its generality. This
new discovery imposes new restrictions on the utility of the BE exciton
transition as a universal \textquotedblleft exciton counter\textquotedblright{}
in experiments on all kinds of semiconductor NCs.

\begin{figure}
\includegraphics[width=1\columnwidth]{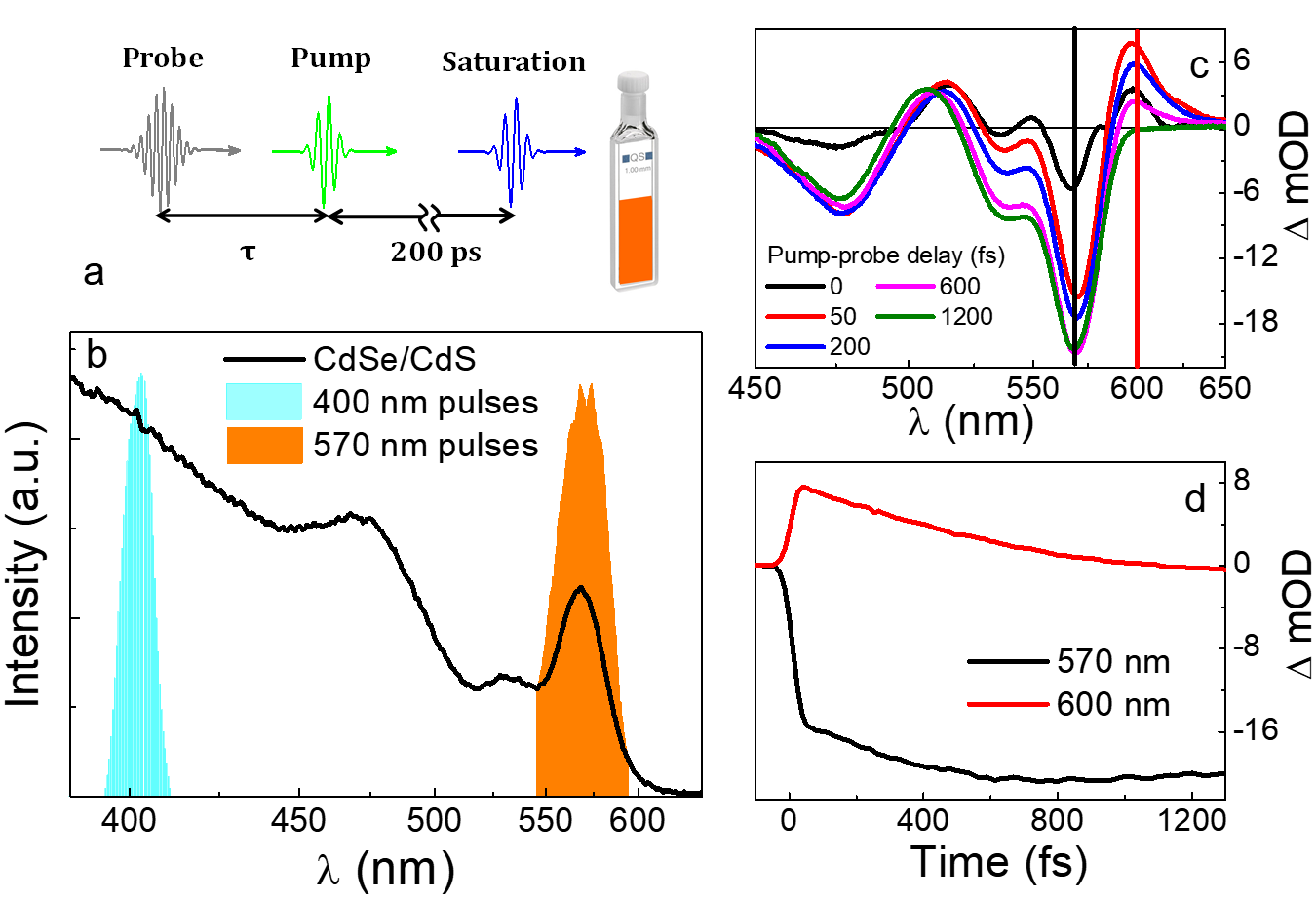}

\caption{\label{fig:Design-of-the}Design of the saturate-pump-probe experiment:
(a) Timing sequence for the 3-pulses. Probe is always a chirped supercontinuum
generated in BaF2. (b) Absorption spectrum of CdSe/CdS NCs dispersed
in hexane, along with the spectra of the excitation pulses (400 and
570 nm). See text for details. c) Overlays of transient absorption
spectra at the designated delays. d) Spectral cuts at the peaks of
the 1S-1S bleach in black, and that of the low energy induced absorption
in red.}
\end{figure}

\section{Results and Discussion}

As shown in Fig.~\ref{fig:Design-of-the}a, cold mono-exciton saturation
is performed by an intense $400nm$ pulse followed by a delay of $200ps$
to allow completion of AR. Fig.~\ref{fig:Design-of-the}b presents
the absorption spectrum of the CdSe/CdS NCs along with the pulse spectra
used for saturation and/or pump pulses in our experiments. Core/shells
were chosen to eliminate the substantial effects of surface trapping
characteristic of bare CdSe cores \citep{Reiss2009,Rabouw2015}. Fig.~\ref{fig:Design-of-the}c
presents an overlay of transient absorption spectra covering the first
2 ps of pump-probe delay. Similar to earlier studies by Kambhampati
and coworkers \citep{Kambhampati2011}, buildup of the 1S-1S PIB is
very rapid, increasing marginally during the subsequent cooling as
depicted in the right panel. The photo-induced absorption band (PIA)
to the red rises at least as rapidly and decays gradually for $\sim2ps$.
These trends are demonstrated by temporal cuts in panel d taken at
wavelengths designated by the color-coded vertical lines in panel
c of Fig.~\ref{fig:Design-of-the}.

\begin{figure}
\begin{centering}
\includegraphics[width=1\columnwidth]{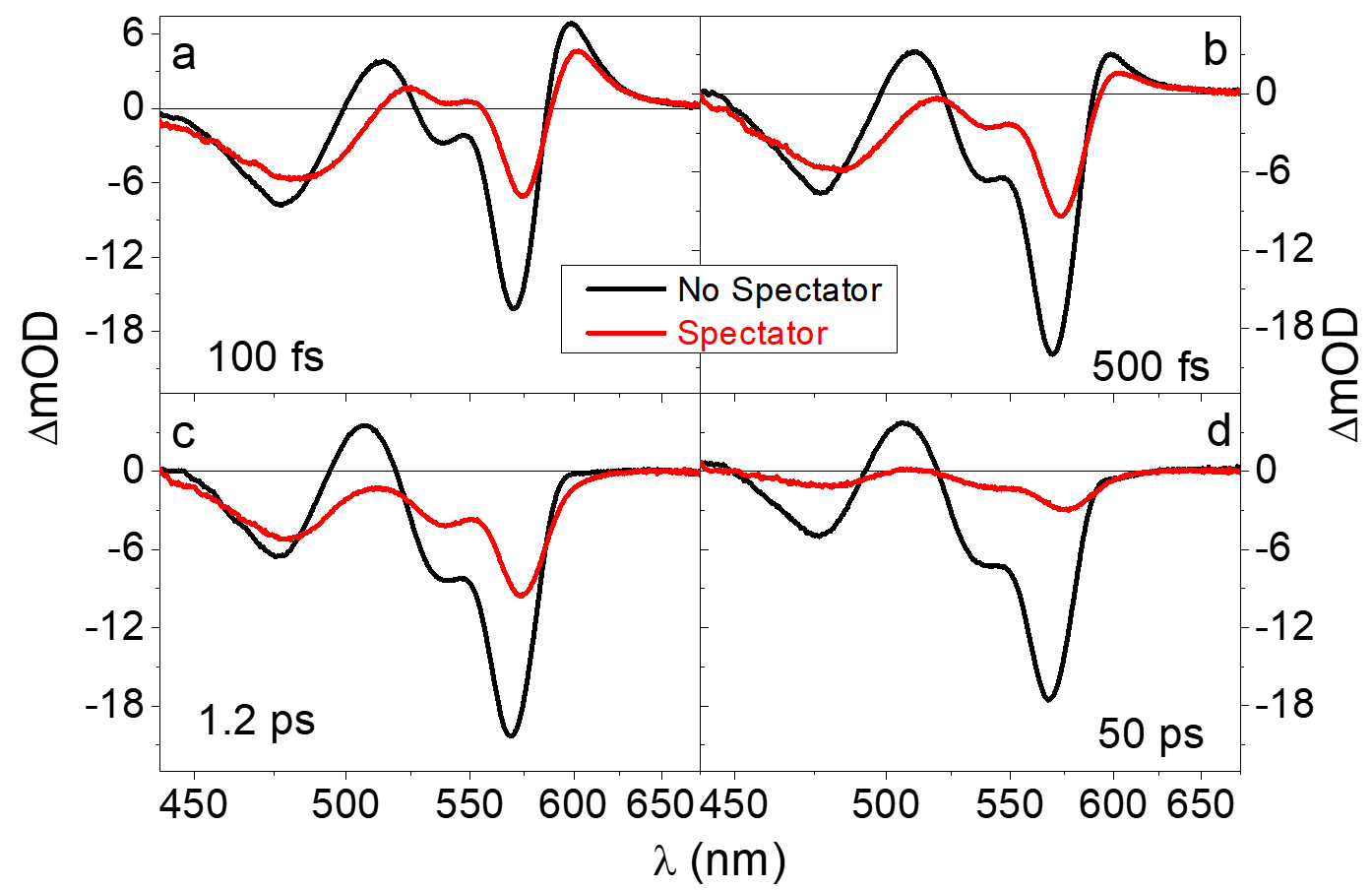}
\par\end{centering}
\caption{\label{fig:Comp-2-3-TA}Comparison of 2- and 3-pulse TA measurements
after weak $400nm$ pumping at a series of designated probe delays
(a-d). Saturation pulses, like the pump, are derived from SHG of the
laser fundamental.}
\end{figure}

Fig.~\ref{fig:Comp-2-3-TA} demonstrates how the presence of a spectator
exciton changes the transient transmission spectra following a femtosecond
$400nm$ excitation. We stress that in both experiments pump fluence
has been controlled to ensure that the probability of absorbing more
than one photon is negligible. Conversely, in the three-pulse experiment
the saturation pulse is much more intense and ensures that $95\%$
of the crystallites absorb at least one photon. The four panels present
transient spectra recorded in both experiments at selected delays
between $100fs$ and $50ps$. Significant differences are apparent
early on, a reduced bleach at the BE peak being the most obvious.
The PIB band in three pulse experiments is also broader and blue shifted
by $6nm$ ($\sim20meV$). Another difference is that the PIA feature
at $520nm$, apparent at all delays in 2-pulse data, is absent once
spectator excitons are present. At 1.2 ps with carrier cooling over,
the remaining PIB associated with state filling is precisely half
as intense in the presence of spectators. These results are in qualitative
agreement with our earlier study on PbSe, but spectator effects in
CdSe are larger. At the $50ps$ delay the amplitude of the BE PIB
with spectators has diminished by $\sim70\%$ due to the presence
of AR which does not affect two pulse pump-probe which involves long
lived single excitons.

\begin{figure}
\begin{centering}
\includegraphics[width=1\columnwidth]{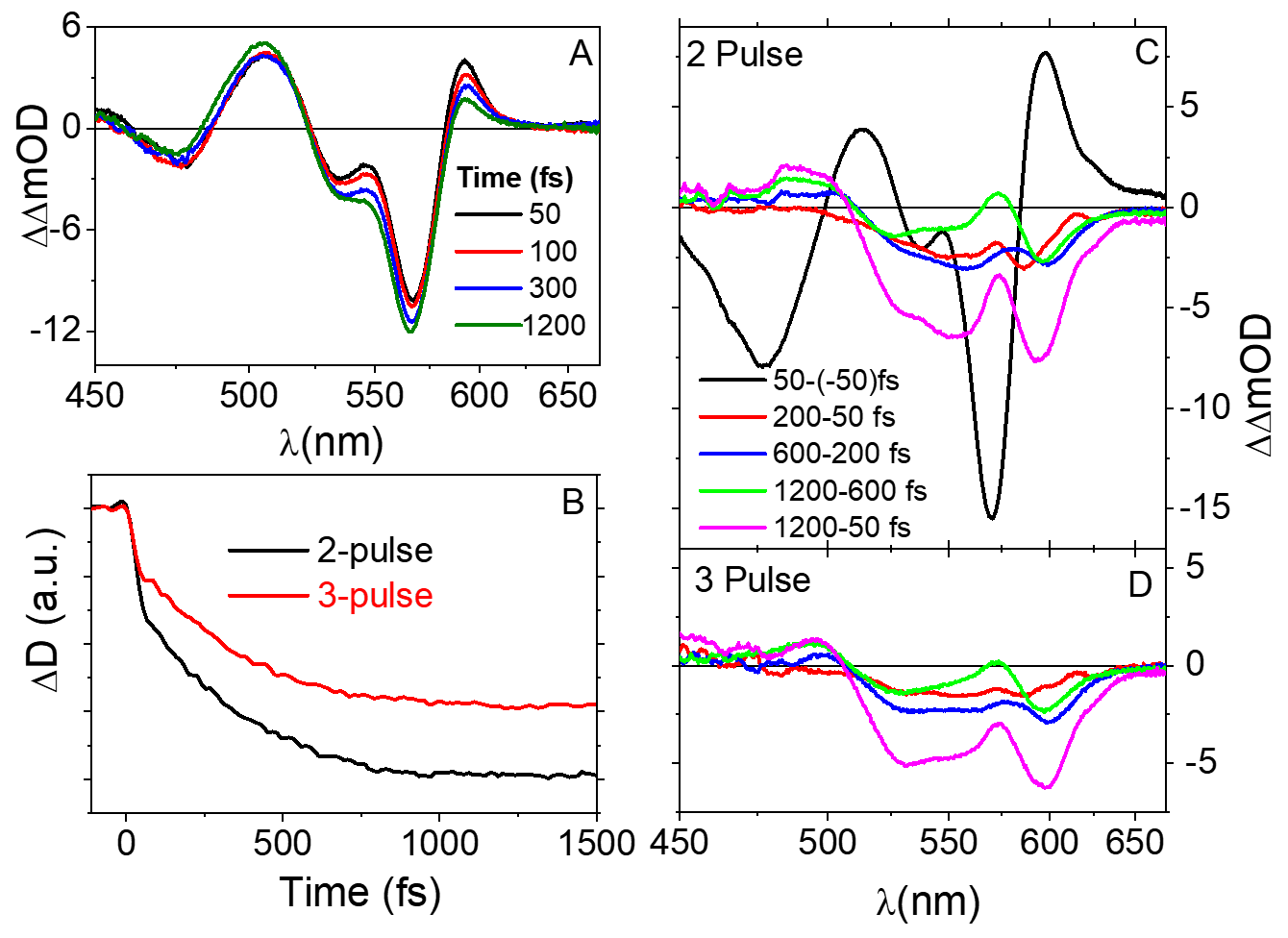}
\par\end{centering}
\caption{\label{fig:Comp-spectral}Comparison of spectral evolution in 2 and
3 pulse experiments pumping at $400nm$. A) Subtraction of the two
data sets at a series of pump-probe delays from $0.05$ to $1.2ps$.
B) Band integrals of the full probed range in the two experiments,
demonstrating the identical cooling kinetics. C) Finite difference
spectra obtained from the 2 pulse data isolating spectral changes
taking place during sequential short intervals of carrier cooling.
D) As in C for three pulse experiments. Notice that these spectra
are essentially identical in both experiments for intervals starting
after $50fs$.}
\end{figure}

A glance at the first three delays in Fig.~\ref{fig:Comp-2-3-TA}
shows that 2- and 3-pulse experiments differ consistently along the
lines detailed above throughout carrier cooling. Notice that while
spectral evolution from delay to delay is pronounced in both experiments,
the difference between 2- and 3-pulse transient spectra at the same
pump-probe delay remains nearly constant. This similarity is demonstrated
in panel A of Fig.~\ref{fig:Comp-spectral} as an overlay of the
subtraction of the pairs presented in Fig.~\ref{fig:Comp-2-3-TA}.
Implications of this are first that carrier cooling dynamics is unaffected
by presence of a BE spectator exciton in this range of delays. This
is further demonstrated by band integrals producing difference dipole
strength over the full spectral range probed in the experiment: $\Delta D\equiv\int_{c/670}^{c/450}\frac{\Delta OD}{\nu}d\nu$,
and plotted in panel B of Fig.~\ref{fig:Comp-spectral}. Panels C
and D display finite difference spectra extracted from 2- and 3-pulse
data, defined as follows:

\begin{align*}
\Delta\Delta OD\left(\lambda,t,\Delta t\right) & =\Delta OD\left(\lambda,t+\frac{\Delta t}{2}\right)\\
 & \,\,\,\,\,-\Delta OD\left(\lambda,t-\frac{\Delta t}{2}\right)
\end{align*}
and serves to isolate spectral changes taking place over a specific
interval of time. These spectra demonstrate essentially identical
spectral evolution taking place in both experiments at delays of $>50fs$.
They also show negligible alterations to the amplitude of PIB after
this point, with spectral changes consisting primarily of the decay
of a broad and structure-less absorption band covering most of the
probed spectra range. A second implication is that dynamic processes
leading to the conserved difference in spectra shown in Fig.~\ref{fig:Comp-spectral}A
must be established considerably before the earliest 50 fs delay.

These consequences will be the subject of a separate report dealing
with carrier cooling dynamics. Here we concentrate on spectra established
after cooling is complete. Clearly one of two assumptions discussed
concerning the bleach amplitude introduced by the second exciton is
wrong. Either the PIB is not linear in the number of excitons, or
else not all excitons directly populate the lowest available conduction
band states. In deciphering this riddle the factor of 2 in BE PIB
between samples with or without resident excitons is a significant
clue. Assume that the PIB is linear in $N$, what could block excitons
from taking their place aside the existing spectator in the $1S^{e}$level?
In order to fill that gap the pair of $1S^{e}$ electrons must be
spin paired, and any disparity in correct orientation could delay
the final step of relaxation. A random distribution of $\left|\uparrow\right\rangle $
and $\left|\downarrow\right\rangle $ spin orientations of relaxing
electrons would prohibit half of the cooling electrons from the BE
until one of the electron spins flip.

\begin{figure}
\begin{centering}
\includegraphics[width=1\columnwidth]{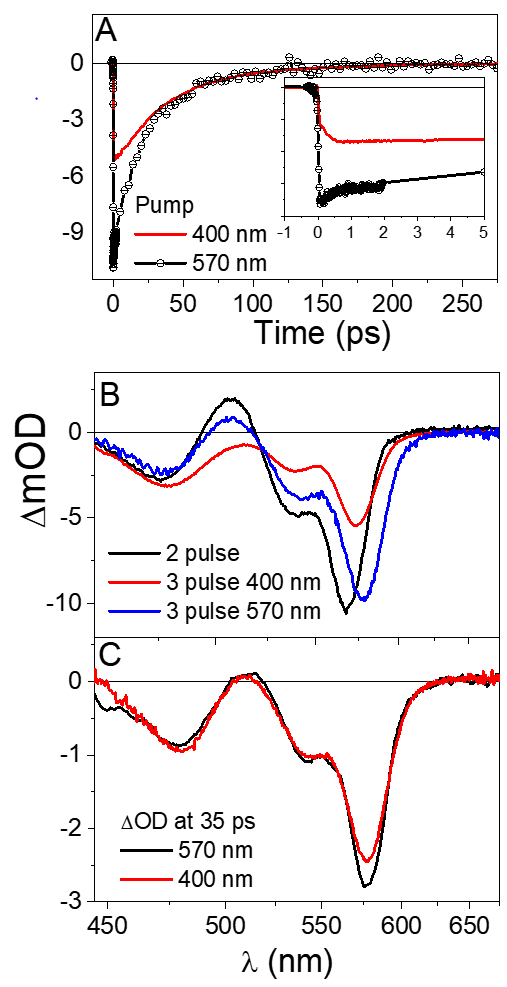}
\par\end{centering}
\caption{\label{fig:Spectral-and-temporal}Spectral and temporal evolutions
from 2-pulse and 3-pulse TA measurements upon pumping at the BE. (A)
Comparison of band edge bleach kinetics for 3-pulse experiments for
different pump excitation conditions: $570nm$ (solid line-circle,
black) or $400nm$ (solid red line). Inset shows expanded scale up
to 5 ps. (B) Comparison of TA spectra at $1.2ps$ for $570nm$ pump
excitation in absence (black) and presence (red) of saturation pulses;
(C) TA spectra of second exciton generated by pumping either at $570nm$
(black) or at 400 nm (red) at a pump-probe delay of $35ps$. Notice
how the initially very different TA spectra of the two experiments
are essentially identical as AR proceeds (See text).}

\end{figure}

Fortunately these assumptions can be tested experimentally. In our
previous study of NC spectator effects pumping was conducted high
in the inter-band continuum since there the cross section for absorption
is unaffected by the existence of other excitons \citep{Spoor2017,Cho2010}.
This provides a trivial method for ensuring an equal dose of additional
excitons deposited in samples with or without spectators. However
a second exciton can also be introduced to spectator containing samples
by exciting directly at the BE. Quantitative comparison of the second
exciton contribution to the PIB requires accounting for the changed
absorption cross section of the pump due to the spectator excitons.
The benefit is that pumping at the BE guarantees that all absorbed
photons populate the $1S^{h}-1S^{e}$ state with or without preexisting
spectator. The results of this experiment are shown in Fig.~\ref{fig:Spectral-and-temporal}.
Panel B presents TA spectra just after cooling for three different
experiments, 2-pulse pump-probe, and two cases of saturate-pump-probe,
the first pumping at $400nm$, and second exciting at 570 nm. Clearly,
the bleach introduced by BE pumping in the presence of spectator excitons
is on par with that apparent in 2-pulse pump-probe. The spectra are
shifted due to the additional bi-exciton interaction involved, and
the absence of residual absorption into the $1S^{h}-1S^{e}$ state
when completely filled. Nonetheless, a band integral demonstrates
that despite these anticipated discrepancies, the notion that BE PIB
varies linearly with the population of the first cold excitons in
the conduction band is upheld.

The alternative is that hot multi-excitons do not all relax to the
lowest energy states. That due to spin orientation conflicts, relaxation
to $1S^{e}$ is partially blocked. This leads to the following predictions:
1) An electron blocked from the BE by virtue of conflicting spin orientation
will be stranded above in the $1P^{e}$ level and induce a partial
PIB of the absorption into this band, and 2) AR kinetics in spin blockaded
bi-excitons will appear to be slower at first since the decay of BE
PIB will be partially canceled by gradual population of $1S^{e}$
following spin flips of the electron (assuming flips take place on
a timescale of \textasciitilde 10 ps). Accordingly, the difference
spectrum in three pulse experiments pumped at 400nm should approach
that obtained by directly exciting into $570nm$ pumping once the
frustrated bi-excitons have annealed to the BE. Our data confirm all
these predictions. In panel B of Fig.~\ref{fig:Spectral-and-temporal},
not only is the PIB by $400nm$ pumping $2$ times smaller than that
induced at $570$, there is a missing PIA feature at $520nm$, consistent
with partial state-filling of $1P^{e}$ due to stranded electrons.
Panel A shows PIB decay kinetics for the same two experiments. As
predicted, the AR dominated decay starts off much more rapidly for
$570$ nm pumping. As the delay increases, both curves merge with
a time constant of $\sim15ps$. If spin frustrated bi-excitons undergo
AR at the same rate as a relaxed one, this is consistent with a spin
flip rate of $\sim1/15ps^{-1}$ per electron (see discussion of spin-flip
mechanism below). Finally, as shown in Fig.~\ref{fig:Spectral-and-temporal}
panel C, the very different TA spectra at $1ps$, asymptotically converge
once the remaining bi-excitons have relaxed to the BE, completing
the consistency test with the spin conflict hypothesis.

\begin{figure}
\begin{centering}
\includegraphics[width=1\columnwidth]{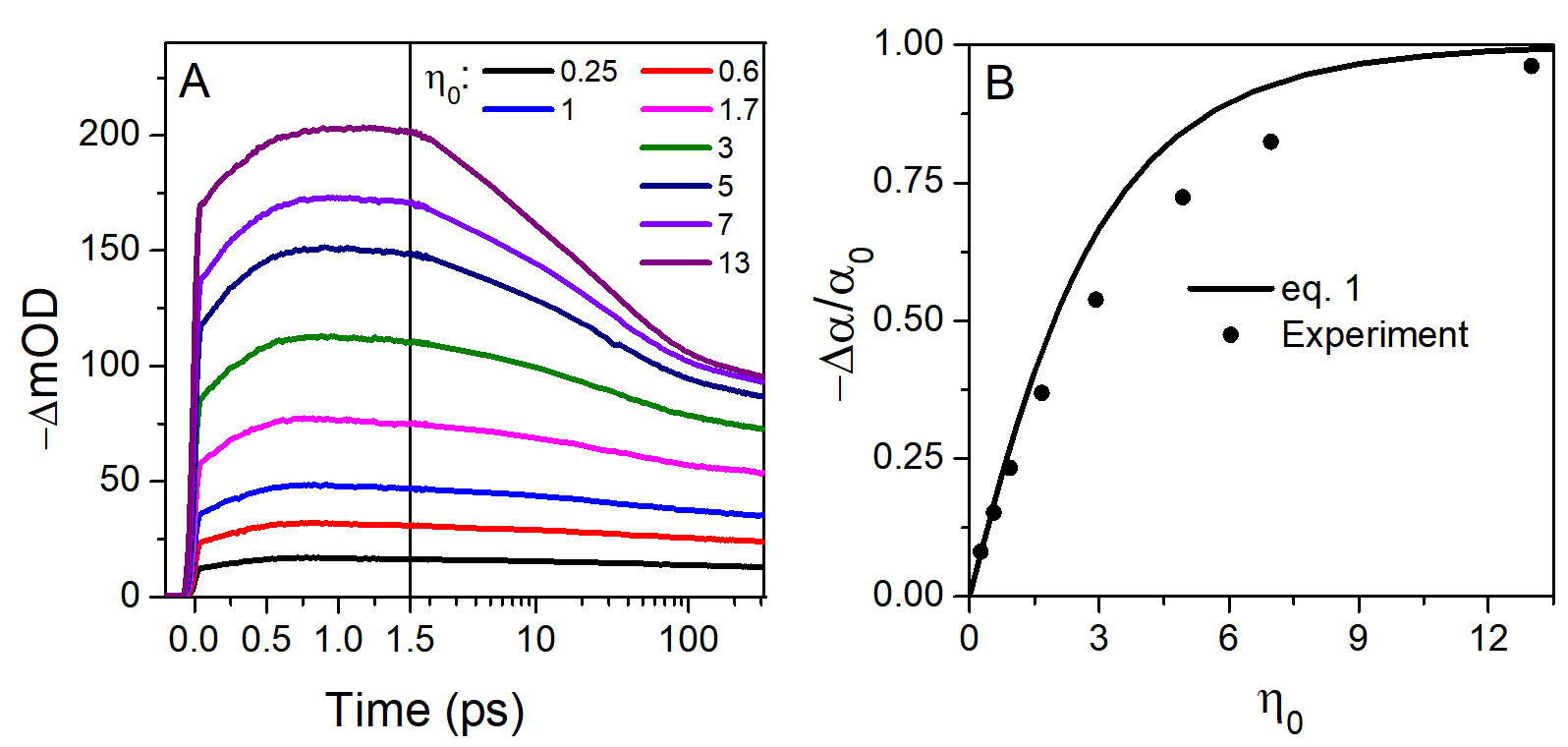}
\par\end{centering}
\caption{\label{fig:BE-Bleach-kinetics}BE Bleach kinetics after $400nm$ excitation
as a function of pump fluence. A) Kinetics from 0-125 ps, on a split
time scale. First $1.5ps$ are plotted on a linear time axis, and
thereafter data is plotted on a logarithmic scale. Pump fluences are
presented in dimensionless units of 0, the average number of absorbed
photons per nanocrystal at the front surface of the sample. B) Comparison
of the measured 1S1S bleach saturation as a function of $\eta_{0}$
compared with a simulated one assuming fast decay of all excitons
to the lowest available states, and state filling of \textonehalf{}
the BE band per exciton.}
\end{figure}

This selection rule for hot multi-exciton relaxation must hold in
any matching scenario. As an example, panel A of Fig.~\ref{fig:BE-Bleach-kinetics}
brings an overlay of pump-probe PIB kinetics for a broad range of
$400nm$ excitation fluences. After determining the absorption cross
section from the bleach amplitude at low pump fluence, as detailed
in the supplementary information, each fluence is designated by $\eta_{0}\equiv\sigma_{A}\times\rho_{h\nu}$,
the average number of absorbed photons per NC at the front surface
of the sample. As $\eta_{0}$ increases, the PIB grows monotonically,
increasing the portion of bleach which decays during AR. As expected,
at the highest pump intensities, this decay accounts for nearly \textonehalf{}
of the bleach maximum at 1.5 ps. Knowing that absorption of the sample
at $400nm$ is unchanged by absorbing even a number of photons, Poisson
statistics integrated throughout the depth of the sample can be used
to calculate the number density of NCs which have absorbed $N$ photons.
Panel B brings the predicted fractional bleach amplitude assuming
that all excitons relax directly to lowest available energy states,
and that each 1S electron bleaches \textonehalf{} of the 1S-1S exciton
band (see details in supplementary information).

As $\eta_{0}$ increases, the actual bleach amplitude falls short
of this prediction. We assign this to the same partially frustrated
relaxation demonstrated in the spectator experiments. As photons are
absorbed, excitons will rapidly relax to the lowest quantized states.
After the first is in place, another will only be able to relax beside
it if it has a correct spin orientation. This situation is worsened
with increased pump fluence as the number of bi-excitons grows. Eventually
as more and more excitons are generated, at least one of the additional
electrons will have a matching spin state to fill the gap and complete
BE bleaching. This explains the ultimate approach of the fractional
to 1 when $\eta_{0}\gg1$. Thus this limitation concerning cooling
of multi-excitons is obvious even in very fundamental pump-probe experiments
once analyzed quantitatively. It must accordingly be considered whenever
the intense band edge exciton bleach is utilized to quantify exciton
occupation numbers when multi-excitons are involved.

To test the plausibility of this interpretation, we developed a 3-state
Lindblad master equation for describing the spin-flip dynamics.\citep{Breuer2002}
The lowest S\textasciicircum e state, well-separated by energy $\varepsilon_{S}$
from $P_{0}^{e}$, is populated by a spin-up spectator electron ($S_{\uparrow}^{e}$).
Another hot electron subsequently promoted by the pump pulse is stranded
in state $P_{-1\text{\textuparrow}}^{e}$, unable to populate $S_{\downarrow}^{e}$
due to Pauli blocking. Further relaxation can be achieved only by
a spin-flip $P_{-1\text{\textuparrow}}^{e}\rightarrow P_{0\text{\textdownarrow}}^{e}$
facilitated by a spin-orbit interaction, after which the decay $P_{0\text{\textdownarrow}}^{e}\text{\textrightarrow}S_{\downarrow}^{e}$
takes place rapidly (Auger cooling). We neglect \textquotedblleft Auger-spin-flip\textquotedblright{}
cooling (where two nearly resonant excitons of opposite spin are Coulomb-coupled:
$\left(nP_{\uparrow}^{h}-P_{\uparrow}^{e}\right)\text{\textrightarrow}\left(mP_{\text{\textuparrow}}^{h}-P_{\text{\textuparrow}}^{e}\right)$
for reasons detailed in the supplementary information). Spin reorientation
due to interaction with nuclear spins is also not included in our
model. Phonon modes serve as a (classical) heat bath for taking up
any energy mismatch. Setting the energy origin at P\_e\textasciicircum 0,
the total Hamiltonian of the system and bath is
\begin{align*}
H & =2\Delta\left|P_{-1\uparrow}^{e}\right\rangle \left\langle P_{-1\uparrow}^{e}\right|-\varepsilon_{S}\left|S_{\downarrow}^{e}\right\rangle \left\langle S_{\downarrow}^{e}\right|\\
 & +s\left(\left|P_{0\downarrow}^{e}\right\rangle \left\langle P_{-1\uparrow}^{e}\right|+\left|P_{-1\uparrow}^{e}\right\rangle \left\langle P_{0\downarrow}^{e}\right|\right)\\
 & +\sum_{j}\left(A_{01}^{j}\left|P_{0\downarrow}^{e}\right\rangle \left\langle S_{\uparrow}^{e}\right|+cc\right)\frac{p_{j}}{m_{j}}\\
 & +\sum_{j}\left(\frac{p_{j}^{2}}{2m_{j}}+\frac{1}{2}m_{j}\omega_{j}^{2}x_{j}^{2}\right)
\end{align*}
Where $-\varepsilon_{S}$ ($2\Delta$) are the energy difference between
the $S^{e}$ ($P_{-1}^{e}$) and the $P_{0}^{e}$ levels respectively,
$s$ is the spin-flip matrix element. For weak electron-nuclear interactions
we neglect all but the linear coupling to the nuclear velocity $p_{j}/m_{j}$.

As a concrete system we considered a $\text{Cd}_{36}\text{Se}_{36}$
cluster with structure taken from Ref.~\onlinecite{gutsev2014structure}
and relaxed using the Q-CHEM density functional (DFT) code \citep{Shao2015}
at the PBE/6-31G level. This gave the following parameter values:
$\varepsilon_{S}=890cm^{-1}$ and $2\Delta=2000cm^{-1}$, while the
spin orbit coupling $s$ was dependent sharply on the identity of
the hole states correlated with the electrons and varied in the range
$0-200cm^{-1}$. Lindblad master equations were set up in accordance
with the Hamiltonian for $T=300K$. $A_{01}$ was determined by requiring
that the $P_{0\downarrow}^{e}\rightarrow S_{\downarrow}^{e}$ decay
time be $0.5ps$ \citep{Guyot-Sionnest1999,Klimov1999,Klimov2000,Sewall2006,Kambhampati2011,Cooney2007a}.
With these parameter values we obtained the temporal populations of
the adiabatic states shown in Figure 6(b). State 2 is of $P_{-1\uparrow}^{e}$
character and at $t=0$ has close to $0.85$ population which decays
through spin-flips, at a rate of about $10ps$, to State 1, which
is of $P_{0\text{\textdownarrow}}^{e}$ character. This latter state
stays slightly populated at later times as it continuously feeds the
S state which is of $S_{\text{\textdownarrow}}^{e}$ character. The
rate of S state population through spin flipping is much longer than
the Auger cooling alone, and takes place on a $10ps$ time-scale.
The sensitivity of these conclusions to the parameters of the model
is rather small, and the fastest rate that can be reached is $\sim4ps$.
These calculations show that a realistic model of the material under
study predicts rates for spin flip limited cooling in the presence
of a band edge spectator which are consistent with that measured experimentally.
This is bolstered by a demonstration of the moderate dependence of
this rate on the material parameters.

\begin{figure}
\includegraphics[width=1\columnwidth]{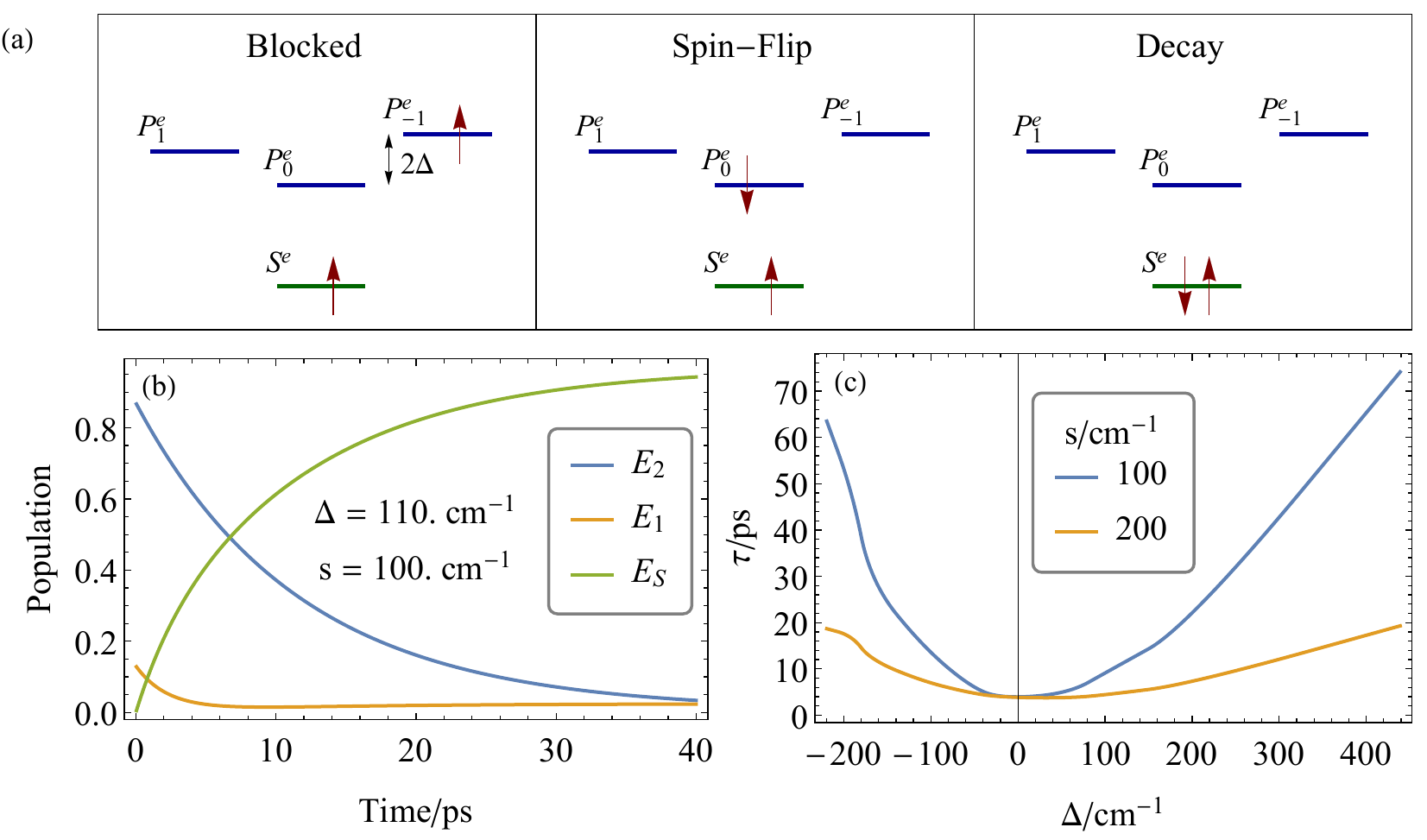}

\caption{\label{fig:(a)-A-schematic}(a) A schematic depiction of the lowest
states at the conduction band-edge, used for modeling the decay rate
of an excited-blocked electron. See text for explanation. (b) Temporal
Lindblad master equation populations for the three adiabatic states
of the model, showing the fast and slow decay transients. (c) Sensitivity
of the long life time $\tau$ to the spin-orbit and level displacement
parameters ($s$ and $\Delta$ respectively) of the model.}
\end{figure}

Relaxation of optically-polarized spin states in nanocrystals has
been investigated extensively, in part to test if semiconductor quantum
dots can provide a platform for spintronics applications \citep{Hanson2008a}.
In the case of colloidal nanocrystals, studies have concentrated on
electronic states limited to the lowest $1S_{3/2}1S^{e}$ exciton
manifold \citep{Gupta1999,Scholes2004,Kim2009}. With spin relaxation
components spanning ps to hundreds of nanosecond timescales, results
of those experiments are not directly comparable to our findings for
a number of reasons. First, the fine structure levels of the exciton
ground state are only defined in terms of correlated hole and electron
microstates, while the observable described here should involve the
electron alone \citep{Klimov1999,Sewall2006}. Furthermore, analysis
of polarization dependent pump-probe experiments conducted in a grating
geometry clarifies that none of the separable phases of spin polarization
decay are dominated by electron spin flips alone \citep{Wong2009}.
Finally, while spin orbit coupling defining the ground exciton sub
levels stems from the \textquotedblleft p\textquotedblright{} orbital
basis of the valence band, in our case its origin is in non-zero orbital
angular momentum related with the envelope $1P^{e}$ function. Our
measurements thus cover a very different process. All that can be
said in comparison is that the observed timescale of a few tens of
ps lies within the broad range characterizing the multi-exponential
decay of optically induced spin polarization in the $1S_{3/2}1S^{e}$
exciton manifold. Only future investigations of how the electron spin
flip rate measured here is effected by temperature, particle size
and crystal morphology will teach more about its relation to other
magnetic relaxation processes in nanocrystals.

\part*{Supplementary Information}

\section{Theoretical model}

We assume The electron states exhibit a separate ground state, $S^{e}$,
$800cm^{-1}$ below a group of three excited electron states, $P_{0,\pm1}^{e}$.
The splitting between the p-states is on the order of $2\Delta=100cm^{-1}$.
We have also performed a TDDFT calculation and estimated various spin-orbit
couplings, between excited states. Because of the multitude of hole
states there is no one number, but a spectrum of couplings, spanning
a range of $s=10-300cm^{-1}$.

\begin{figure}
\begin{centering}
\includegraphics[width=1\columnwidth]{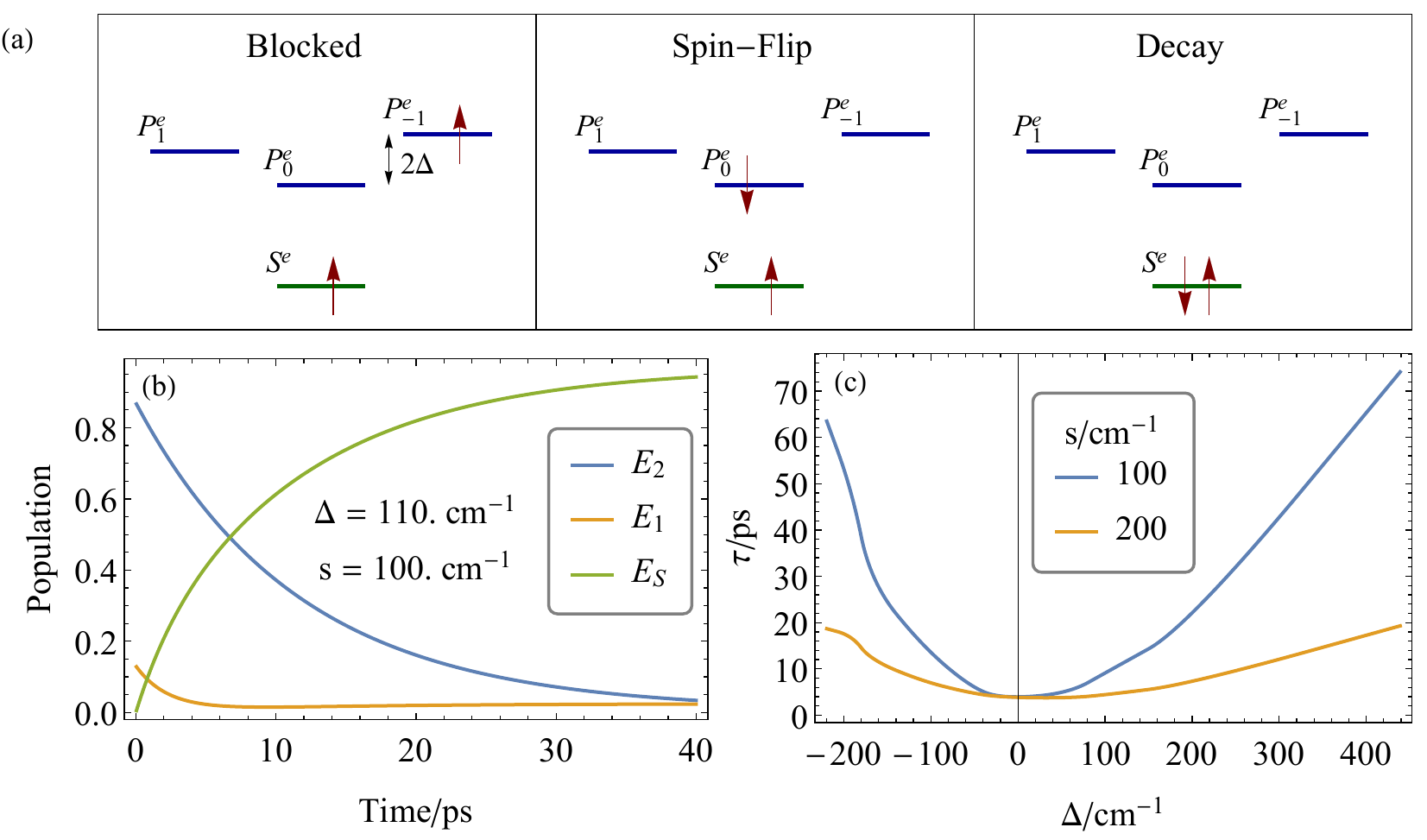}
\par\end{centering}
\caption{\label{fig:states}(a) Left panel: the up-spin electron in the $P_{-1\uparrow}^{e}$
level (say) is \emph{blocked} from decay into the edge level $S_{\downarrow}^{e}$
due to Pauli exclusion. Middle panel: A spin-orbit induced \emph{spin-flip}
occurs: $P_{-1\uparrow}^{e}\to P_{0\downarrow}^{e}$. Right panel:
The down-spin electron decay $P_{0\downarrow}^{e}\to S_{\downarrow}^{e}$.
(b) The populations of the three adiabatic states as a function of
time for the indicated values of $\Delta$ and $s$. (c) a sensitivity
analysis checking time scale for the final cooling to the parameters.}
\end{figure}

In Fig.~\ref{fig:states} we describe our model for the experiment
in CdSe NCs. There are two excitons, one is ``cold'', in the band
edge, due to the first experimental pulse with an up-spin electron
in the $S^{e}$ state and the second has its hole at the top of the
valence manifold but its up-spin electron is in one of the P states
of the and it is blocked from further decay onto the $S^{e}$ level
because of Pauli exclusion. 

\subsection{Dissipative dynamics under spin-flip}

We first neglect spin-orbit coupling. In this case, we consider the
electronic Hamiltonian of the levels $P_{-1\uparrow}^{e}$, $P_{0\downarrow}^{e}$
and $S_{\downarrow}^{e}$: 
\begin{align}
H_{S} & =\left(\begin{array}{ccc}
2\Delta & 0 & 0\\
0 & 0 & 0\\
0 & 0 & -\varepsilon_{S}
\end{array}\right).\label{eq:HS-I}
\end{align}
where we assumed that the energy of the $P_{0\downarrow}^{e}$ level
is zero and that of the $P_{-1\uparrow}^{e}$ level is displaced by
$2\Delta$ while that of the $S^{e}$ is below by $\varepsilon_{S}>0$. 

The total Hamiltonian $H_{tot}$ describing the coupling of to a bath
of phonons is
\begin{align}
H_{tot} & =H_{B}+H_{S}+\sum_{j}\left(\frac{p_{j}}{m_{j}}\right)\left(\begin{array}{ccc}
0 & 0 & 0\\
0 & 0 & A_{j}^{12}\\
0 & A_{j}^{21} & 0
\end{array}\right)
\end{align}
where $A_{j}^{12}=A_{j}^{21*}$ are the non-adiabatic coupling terms
and 
\begin{equation}
H_{B}=\sum_{j}\left(\frac{p_{j}^{2}}{2m_{j}}+\frac{1}{2}m_{j}\omega_{j}^{2}x_{j}^{2}\right)
\end{equation}
is the phonon Hamiltonian. Notice that there is no non-adiabatic coupling
of the $P_{-1\uparrow}^{e}$ level and the $S_{\downarrow}^{e}$ level
because of the different spins. 

Now, introduce the spin coupling $s$ between levels $P_{-1\uparrow}^{e}$
and $P_{0\downarrow}^{e}$. The 3-level Hamiltonian in Eq.~\ref{eq:HS-I}
is thus generalized to 
\begin{align}
H_{S} & =\left(\begin{array}{ccc}
2\Delta & s & 0\\
s & 0 & 0\\
0 & 0 & -\varepsilon_{S}
\end{array}\right).\label{eq:HS-II}
\end{align}
We diagonalize $H_{a}=U^{\dagger}H_{S}U$, using 
\begin{align}
U & =\left(\begin{array}{ccc}
\cos\theta/2 & -\sin\theta/2 & 0\\
\sin\theta/2 & \cos\theta/2 & 0\\
0 & 0 & 1
\end{array}\right)
\end{align}
giving:
\begin{align}
H_{a} & \equiv\left(\begin{array}{ccc}
E_{1} & 0 & 0\\
0 & E_{2} & 0\\
0 & 0 & E_{S}
\end{array}\right)\\
 & =2R\left(\begin{array}{ccc}
\cos^{2}\theta/2 & 0 & 0\\
0 & -\sin^{2}\theta/2 & 0\\
0 & 0 & -\varepsilon_{S}/2R
\end{array}\right)
\end{align}
where, $R$ and $\theta$ are defined by requiring $\Delta=R\cos\theta$
and $s=R\sin\theta$. The total Hamiltonian then becomes:
\begin{equation}
H_{tot}^{a}=H_{a}+H_{B}+\sum_{j}\left(\frac{p_{j}}{m_{j}}\right)\left(\begin{array}{ccc}
0 & 0 & \tilde{A}_{j}^{01}\\
0 & 0 & \tilde{A}_{j}^{12}\\
\tilde{A}_{j}^{10} & \tilde{A}_{j}^{21} & 0
\end{array}\right).
\end{equation}
where $\tilde{A}_{j}^{12}=\tilde{A}_{j}^{21*}=A_{j}^{12}\cos\theta/2$
and $\tilde{A}_{j}^{02}=\tilde{A}_{j}^{20*}=A_{j}^{12}\sin\theta/2$
now couple both adiabatic P-states with the S-state because the spin-orbit
coupling mixes the up- and down-spin states. 

In order to study the dynamics, we assume the reduced $3\times3$
density matrix $\sigma_{ij}$ obeys the Lindblad equation\citep{Breuer2002}:
\begin{equation}
\dot{\sigma}\left(t\right)=-\frac{i}{\hbar}\left[H_{a}+H_{1a},\sigma\right]+\text{\ensuremath{\mathfrak{D}}}\sigma\left(t\right)\label{eq:Lindblad}
\end{equation}
where the dissipative and hermitian bath effects are given as:
\begin{align}
\text{\ensuremath{\mathfrak{D}}}\sigma & =\sum_{\omega}\gamma\left(\omega\right)\left[L\left(\omega\right)\sigma L\left(\omega\right)^{\dagger}-\frac{1}{2}\left\{ L\left(\omega\right)^{\dagger}L\left(\omega\right),\sigma\right\} \right]\label{eq:DissipatorGen}\\
H_{1} & =\sum_{\omega}S\left(\omega\right)L\left(\omega\right)^{\dagger}L\left(\omega\right)\label{eq:H1Gen}
\end{align}
where the summation is over discrete frequencies, $\omega_{m}=m\frac{2\pi}{T_{d}}$
where $T_{d}$ is the discretization period and
\begin{equation}
\Gamma\left(\omega\right)=\frac{1}{2}\gamma\left(\omega\right)+iS\left(\omega\right)=\int_{0}^{T_{d}}e^{i\omega t}\left\langle \frac{p\left(t\right)}{m}\frac{p\left(0\right)}{m}\right\rangle _{\beta}dt\label{eq:bathResponse}
\end{equation}
is the bath frequency velocity autocorrelation function at inverse
temperature $\beta$ (see Appendix~\ref{sec:The-Harmonic-correlation}).
For the two resonant frequencies, $\hbar\omega_{20}=\varepsilon_{S}+2R\cos^{2}\theta/2$
and $\hbar\omega_{21}=\varepsilon_{S}-2R\sin^{2}\theta/2$ and present
in our Hamiltonian, we have the following Lindblad operator

\begin{align}
L_{\omega} & =\hbar^{-1}\left(\begin{array}{ccc}
0 & 0 & \tilde{A}^{02}\delta_{\omega,\omega_{20}}\\
0 & 0 & \tilde{A}_{j}^{12}\delta_{\omega,\omega_{21}}\\
\tilde{A}_{j}^{20}\delta_{-\omega,\omega_{20}} & \tilde{A}_{j}^{21}\delta_{-\omega,\omega_{21}} & 0
\end{array}\right)
\end{align}
which is a Liouvillian eigenstate $\left[H,L_{\omega}\right]=\hbar\omega L_{\omega}$.
With this, the dissipation above term in (\ref{eq:DissipatorGen})
becomes \onecolumngrid

\begin{align}
\text{\ensuremath{\mathfrak{D}}}\sigma & =\left(\begin{array}{ccc}
-D_{20}\sigma_{00}+D_{02}\sigma_{22} & -\frac{D_{20}+D_{21}}{2}\sigma_{01} & -\frac{D_{20}+D_{02}+D_{12}}{2}\sigma_{02}\\
-\frac{D_{20}+D_{21}}{2}\sigma_{10} & -D_{21}\sigma_{11}+D_{12}\sigma_{22} & -\frac{D_{02}+D_{12}+D_{21}}{2}\sigma_{12}\\
-\frac{D_{20}+D_{02}+D_{12}}{2}\sigma_{20} & -\frac{D_{02}+D_{12}+D_{21}}{2}\sigma_{21} & -\left(D_{02}+D_{12}\right)\sigma_{22}+D_{20}\sigma_{00}+D_{21}\sigma_{11}
\end{array}\right),\label{eq:dissipator}
\end{align}
\twocolumngrid

where (see Appendix\_\ref{sec:The-Harmonic-correlation}, Eq.~(\ref{eq:gamma})):
\begin{align}
D_{02} & =a\omega_{02}\left(\left\langle n\right\rangle _{\beta\ensuremath{\hbar}\omega_{02}}+1\right)\sin^{2}\theta/2,\\
D_{20} & =a\omega_{02}\left\langle n\right\rangle _{\beta\ensuremath{\hbar}\omega_{02}}\sin^{2}\theta/2,
\end{align}

\begin{align}
D_{12} & =a\omega_{12}\left(\left\langle n\right\rangle _{\beta\ensuremath{\hbar}\omega_{12}}+1\right)\cos^{2}\theta/2,\\
D_{21} & =a\omega_{12}\left\langle n\right\rangle _{\beta\ensuremath{\hbar}\omega_{12}}\cos^{2}\theta/2.
\end{align}
with $a=T_{d}\frac{\left|A_{12}\right|^{2}}{2m\hbar}$ the dimensionless
non-adiabatic parameter. 

\subsection{Why spin-flip Auger coupling is weak}

In CdSe NCs Auger coupling is a major nonradiative energy dissipation
mechanism in excited NCs \citep{Klimov2000b}. The direct coupling
between excitons involves the matrix elements :
\begin{align}
\left\langle 0\left|c_{j\uparrow}^{\dagger}c_{b\uparrow}\hat{V}c_{a\uparrow}^{\dagger}c_{i\uparrow}\right|0\right\rangle  & =\delta_{ij}\delta_{ab}2\left(2V_{vssv}-V_{svsv}\right)n_{s}n_{v}\nonumber \\
 & +2\delta_{ij}\left(2V_{btta}-V_{btat}\right)n_{t}\nonumber \\
 & -2\delta_{ab}\left(2V_{ittj}-V_{itjt}\right)n_{t}\label{eq:direct-auger}\\
 & +2\left(V_{ibja}-V_{ibaj}\right)\nonumber 
\end{align}
where $\hat{V}=\sum_{stuv}\sum_{\sigma\sigma'=\uparrow\downarrow}V_{utsv}c_{u\sigma}^{\dagger}c_{t\sigma'}^{\dagger}c_{s\sigma'}c_{v\sigma}$
is the Coulomb interaction in second quantization, $\psi_{s}\left(\boldsymbol{r}\right)$,
$\psi_{t}\left(\boldsymbol{r}\right)$, $\psi_{u}\left(\boldsymbol{r}\right)$,
$\psi_{v}\left(\boldsymbol{r}\right)$ are one-electron (Hartree-Fock)
eigenstates and 
\begin{align}
V_{utsv} & =\frac{1}{2}\iint\frac{\left[\psi_{u}\left(\boldsymbol{r}'\right)\psi_{v}\left(\boldsymbol{r}'\right)\right]\left[\psi_{t}\left(\boldsymbol{r}\right)\psi_{s}\left(\boldsymbol{r}\right)\right]}{\left|\boldsymbol{r}-\boldsymbol{r}'\right|}d\boldsymbol{r}d\boldsymbol{r}'
\end{align}
and for which the creation and annihilation commutation relations
are:
\begin{equation}
\left[c_{u\sigma}^{\dagger},c_{s\sigma'}\right]_{+}=\delta_{\sigma\sigma'}\delta_{uv}
\end{equation}
 The first term in Eq.~(\ref{eq:direct-auger}), $i=j$ and $a=b$
is a ``diagonal element'' and is not interesting for our purpose.
When \emph{either }$a=b$ \emph{or }$i=j$ (but not both) it's the
second (or third) term that is important. This is either that the
electron changes state while the hole is a spectator or vice versa.
Suppose it's the electron changing from $a$ to $b$ while the hole
stays put in $i=j$, the coupling then involves the Coulomb interaction
of a electron charge distribution $2\psi_{b}\left(\boldsymbol{r}\right)\psi_{a}\left(\boldsymbol{r}\right)$
with the \emph{entire }electron density 2$\sum_{i=1}^{N_{e}/2}\left|\psi_{i}\left(\boldsymbol{r}\right)\right|^{2}$
in the nanocrystals (corrected by a much smaller exchange term, since
only $\psi_{t}$'s that overlap with both $\psi_{a}$ and $\psi_{b}$
contribute). So, as long as $\psi_{a}\left(\boldsymbol{r}\right)$
and $\psi_{b}\left(\boldsymbol{r}\right)$ strongly overlap in space
this is very strong. 

However, if the coupling involves spin-flip, the corresponding Auger
element evaluates just one term
\begin{equation}
\left\langle 0\left|c_{j\downarrow}^{\dagger}c_{b\downarrow}\hat{V}c_{a\uparrow}^{\dagger}c_{i\uparrow}\right|0\right\rangle =2V_{biaj}\label{eq:sf-auger}
\end{equation}
which is the Coulomb interaction between two charge of distributions
one is $\psi_{b}\left(\boldsymbol{r}\right)\psi_{j}\left(\boldsymbol{r}\right)$
and the other $\psi_{i}\left(\boldsymbol{r}\right)\psi_{j}\left(\boldsymbol{r}\right)$,
for this to be strong one requirement is good electron-hole overlap
for each of the two excitons and that they are reasonably close so
that the Coulomb interaction is considerable. The direct Auger involves
interacting of the dynamical electron with all other electrons in
the system while the spin-flip process is just a local two-electron
interaction.

Summarizing, Auger processes which are typically efficient mechanisms
for biexciton decay in nanocrystals are much slower when a spin-flip
is involved. 

\subsection{Results of model}

The model involves 5 parameters: the temperature $T$, taken to be
300K, energy offset of the $S_{\downarrow}^{e}$ state $\varepsilon_{S}$
with respect to the $P_{e}^{0}$ state, the dimensionless non-adiabatic
coupling strength $a$, the P-level splitting $\Delta$ and the spin-flip
coupling strength $s$. 

Using Q-CHEM \citep{Shao2015}) we have performed \emph{ab initio
}calculations on relaxed $\text{Cd}_{36}\text{Se}_{36}$ (see Ref.~\onlinecite{gutsev2014structure}).
We used the PW91 functional and a small basis set (see a separate
supplementary information file). The calculations reveal the typical
CdSe NC frontier orbital structure: a dense manifold of hole states
and a sparse one for electron states above it, with an optical gap,
of 1.1eV. We also determined from the calculation $\varepsilon_{S}=0.1eV$
and $\Delta=500\mu E_{h}$. Finally we determined the parameter $a=0.14$,
by requiring that the rate of decay $P\to S$ without a spectator
(without the requirement for spin-flip) be $0.5ps$, in accordance
with the known experiment estimates. 

The spin-flip mechanism considered here is caused by the spin-orbit
coupling between a pair of singlet and triplet excitons. The singlet
exciton, originally formed by the laser pulse, has decayed in a fast
time scale to the blocked state, where the electron is in a $P^{e}$
excited state with the spin. The triplet exciton is similar except
that the excited $P^{e}$ electron has flipped its spin (and changed
to a different $P^{e}$ state, see Fig.~\ref{fig:states}a). Due
to the large density of hole states in our model $\text{Cd}_{36}\text{Se}_{36}$
system, we were forced to produce a large number of exciton states:
the electron in the first 35 excitons was in the lowest $S^{e}$ state
and only the hole was in an excited state and only in exciton number
36, at energy of 1.34eV, was the first to have an electron in an excited
$P^{e}$ state . Out of the first 200 excitons we selected the few
that were 1) dominated by a single electron-hole excitation, and 2)
the electron was in one of the two lowest diabatic $P^{e}$ states.
Due to non-adiabatic effects these states mix slightly and we label
the adiabatic states as $1$ (mostly $P_{0}^{e}$) and $2$ (mostly
$P_{-1}^{e}$). We focused attention on a small energy band 1.34-1.56eV
which contained about 100 excitons (out of the 200 calculated). Of
those, only nine were singlet excitons, four with the electron in
state 1 and five with the electron in state $2$. We also identified
sixteen such triplet excitons, nine having an electron in state $1$
and seven an electron in state $2$. Within these excitons there are
four types of SO couplings: $\left\langle S_{1}\left|H_{SO}\right|T_{1}\right\rangle $,
$\left\langle S_{2}\left|H_{SO}\right|T_{2}\right\rangle $, $\left\langle S_{1}\left|H_{SO}\right|S_{2}\right\rangle $
and $\left\langle S_{2}\left|H_{SO}\right|S_{1}\right\rangle $. The
first pair of matrix elements describe a hole spin-flip (the electron
stays put) while the second pair describe an electron spin-flip. The
calculations show, that the hole spin-flip (which is not of interest
for our mechanism) involved a strong coupling averaging at $200cm^{-1}$
and peaking at $480cm^{-1}$. On the other hand, \emph{electron }spin-flip
SO couplings typically have 5 times weaker couplings: averaging at
$50cm^{-1}$ and peaking at $144cm^{-1}$. These calculated \emph{ab
initio }values prompted us to take a representative value of $s=100cm^{-1}$
and then also comparing to $s=200cm^{-1}$as a stability analysis. 

With these parameters we solved the Lindblad equation and taking (see
Fig.~\ref{fig:states}(a)) and obtained the adiabatic state populations
$\sigma_{ii}\left(t\right)$, $i=2,1,S^{e}$. The system starts in
(diabatic) state $P_{-1}^{e}$ which populates mainly state 2 and
due to spin-orbit coupling also a bit of state 2 (this small part
fasyer to the $S^{e}$ state). Then the main part of the population
decays at a time scale of 10ps, reaching equilibrium.

We also checked the sensitivity of the results to $s$ and $\Delta$
and obtained the lifetime in Fig.\ref{fig:states}(c). It is interesting
to see that for large spin-orbit couplings even 200$cm^{-1}$ the
spin-flip effect slow down to about 5 ps.

\section{Materials and Methods}

\subsection{Synthesis of CdSe QDs core}

A mixture of cadmium oxide (CdO) and oleic acid (OA) in a molar ratio
of 1:4 and 7.5 mL of 1-octadecene (ODE) was put in a 25 mL three-neck
flask. The reaction mixture was degassed for 1 h at $100^{o}C$ under
vacuum. Under nitrogen, the temperature was then raised to $300^{o}C$
until the solution turned clear, indicating the formation of cadmium
oleate. Then the solution was cooled, and the octadecylamine (ODA)
was added in a molar ratio of 1:8 (Cd/ODA). Afterward, the solution
was heated to $280^{o}C$ , and 8 mL of 0.25 M trioctylphosphine selenide
(TOPSe) was injected under vigorous stirring. The growth was terminated
after 16 minute by rapid injection of 10 ml of ODE and the reaction
mixture was further cooled down by water bath. As prepared core CdSe
QDs were precipitated twice with a 2-propanol/ethanol mixture (1:1\textminus 1:2),
separated by centrifugation, and redissolved in hexane. 

\subsection{Synthesis of CdSe QDs coated with 1 monolayer (ML) of CdS}

The 0.1 M Cd precursor was prepared by dissolving 0.1 mmol of cadmium
acetate ($Cd(Ac)_{2}$) and 0.2 mmol of hexadecylamine (HAD) in 10
mL of ODE at $100^{o}C$ inside a nitrogen-filled glovebox. The 0.1
M S precursor was prepared by dissolving 0.1 mmol of sulfur in 10
mL of ODE at $100^{o}C$ . The coating procedure has been adapted
from previous report \citep{Grumbach2015}. Initially, 14.8 mL of
ODE was degassed under vacuum for 1 h at $100^{o}C$. The ODE was
cooled to $65^{o}C$, and a solution of 7.4 \texttimes{} 10\textminus 4
mmol CdSe QDs in hexane was injected. Then, the Cd precursor solution
was added at $65^{o}C$. After degassing for 10 min under vacuum at
$65^{o}C$ , the S precursor was added. The reaction mixture was quickly
heated to $100^{o}C$ and allowed to remain for 2 h at this temperature.
Then the nanoparticles were precipitated with a mixture of 2-propanol
and ethanol (1:1\textminus 1:2), separated by centrifugation, and
re-dissolved in hexane. 

\subsection{Pump-probe measurements}

The CdSe/CdS core/shell QDs were prepared under inert atmosphere,
inside a nitrogen filled glove box. The sample was placed in air-tight
0.25 mm quartz cell for the pump-probe measurements. A home built
multi-pass amplified Ti-Sapphire laser producing 30 fs pulses at 790
nm with 1 mJ of energy at 1 kHz repetition rate was used to generate
the fundamental. The laser fundamental was split into different paths
for generation of probe and pump pulses. The pump pulses at 400 nm
were produced by frequency doubling of the fundamental 800 nm pulses,
whereas the pump pulses at 570 nm were generated by second harmonic
generation of signal (at 1140 nm) from an optical parametric amplifier
(TOPAS 800, Light Conversion). The pump pulses at 570 nm were compressed
using grating-mirror compressor set up. The spectra of two pump pulses
are shown in Figure 1b. The supercontinuum probe pulses were generated
by focusing 1200 nm output pulses of another optical parametric amplifier
(TOPAS Prime, Light Conversion) on a 2 mm BaF2 crystal. The pump and
probe beams were directed to the sample using all reflective optics.
The spot size of the pump on the sample was at least two times larger
compared to that of the probe beam. 

Conventional two-pulse pump-probe experiments were carried out with
white light ranging from 420-700 nm as probe and 400 nm or 570 nm
pulses as pump, with low fluence such that the average no. of exciton
per nanocrystals, \textgreek{h} \ensuremath{\approx} 0.2. In the case
of three-pulse measurements, the same two-pulse pump-probe experiments
were repeated in presence of another saturation pulse at 400 nm which
arrives \textasciitilde 200 ps earlier than the pump pulses. Raw
data from three-pulse measurements show a constant bleach (\textasciitilde 10\%
that of initial single exciton bleach) signal even long time after
completion of Auger recombination, suggesting that \textasciitilde 10\%
of the total no. of NCs were not saturated by the strong saturation
pulses. Thus, raw data from three-pulse measurements was first subtracted
by 10\% of two-pulse data and then the subtracted dataset was divided
by 0.9 so that the final three-pulse data represents for fully saturated
sample. 

\section{Experimental determination of absorption cross section of the CdSe/CdS
core/shell nanocrystals}

\begin{figure}
\begin{centering}
\includegraphics[width=1\columnwidth]{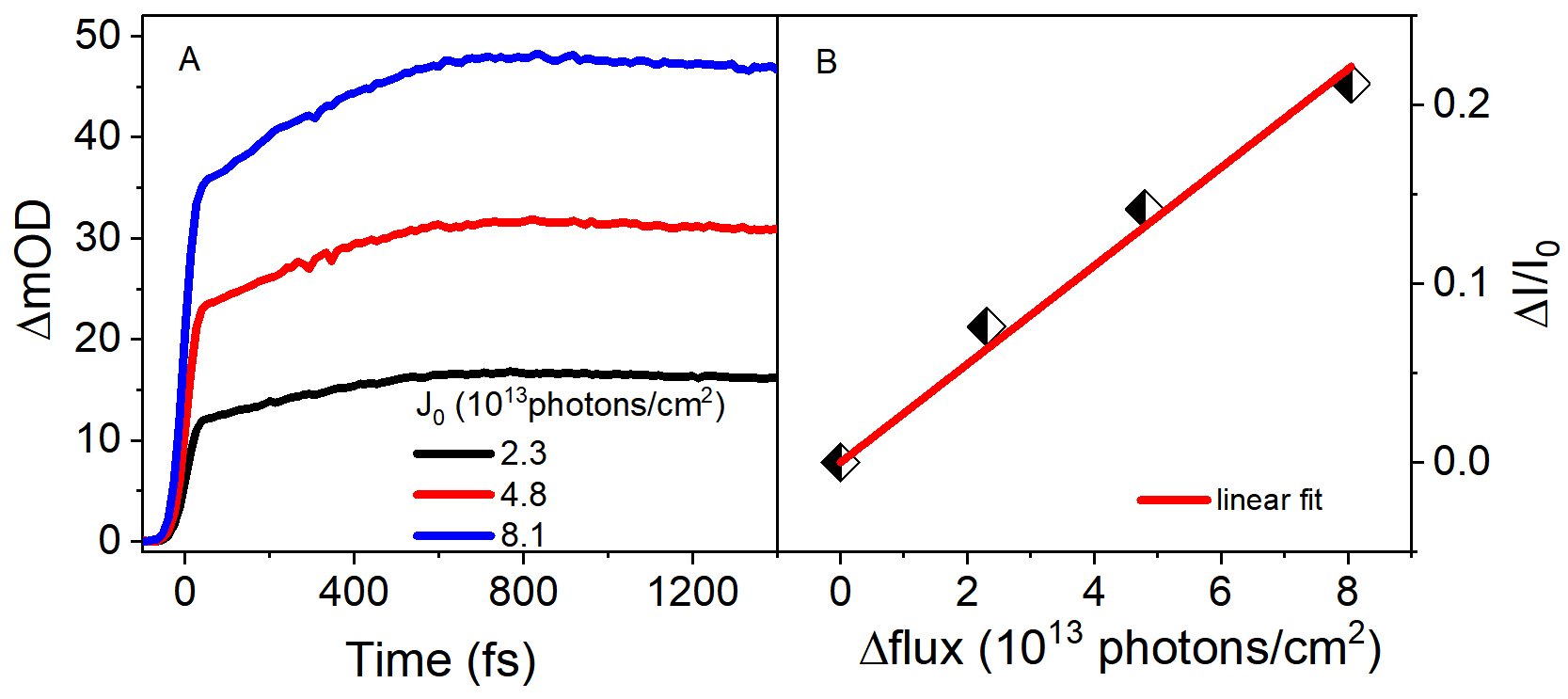}
\par\end{centering}
\caption{\label{fig:Determination-of-absorption}Determination of absorption
cross section of CdSe/CdS core/shell nanocrystals. (A) Plot of 1S1S
bleach signal (at 570 nm) vs. pump-probe delay (up to 1.4 ps) after
excitation with 400 nm pulses, at different excitation pump fluences
as presented in the inset; (B) Plot of fractional absorption change
vs. photon flux and linear fit of the experimental data. The absorption
cross section can be calculated from the slope of this fit.}

\end{figure}

The absorption cross section of the CdSe/CdS core/shell QDs was experimentally
determined according to previously reported procedure \citep{Gdor2015}.
Fig.~\ref{fig:Determination-of-absorption}A demonstrates the 1S1S
transition bleach as a function of pump-probe delay after $400nm$
excitation, for a series of different excitation photon fluxes as
indicated in the inset. All these experiments were carried out with
weak excitation fluences such that the bleach changes linearly. Fig.~\ref{fig:Determination-of-absorption}B
shows a plot of \ensuremath{\mathit{\Delta}}I/I0 vs. photon flux change.
From the linear fit of these experimental data points and known the
2-fold degeneracy of 1S1S transition of the CdSe dots, the slope of
the fit gives the absorption cross section value at the pump wavelength,
$\sigma_{400}=(2.7\pm0.1)\times10^{-15}cm^{2}$.

\section{Simulation of excitation pump fluence dependence on band edge bleach}

Defining $J_{0}$ and $J_{\infty}$ as the pump photon flux (photons/$cm^{2}$)
in the front and back surfaces of an optically thick sample cell,
and $\sigma$ as the absorption cross-section ($cm^{2}$) at the pump
wavelength, the average no. of absorbed photons per nanocrystal at
the front ($\eta_{0}$) and back surface ($\eta_{\infty}$) are:
\begin{equation}
J_{0}\sigma=\eta_{0},\,\,\,\,\,J_{\infty}\sigma=\eta_{\infty}.
\end{equation}
Denoting $\rho_{T}$ as the total density/$cm^{2}$ of NC in the sample,
assuming pump wavelength absorption is unaffected by existing excitons
the ratio $\eta_{\infty}/\eta_{0}$ must equal to $e^{-\rho_{T}\sigma}$
and therefore:
\begin{align}
\rho_{T} & =\frac{1}{\sigma}\log\left(\frac{\eta_{0}}{\eta_{\infty}}\right)\label{eq:rho_T}\\
 & =\frac{1}{\sigma}\int_{\eta_{\infty}}^{\eta_{0}}\frac{d\eta}{\eta}.\nonumber 
\end{align}
For an optically thin slab of nanodots at a given $\eta$ the probability
for a quantum dot to absorb $N$ pump photons follows the Poisson
distribution,
\begin{equation}
P_{N}\left(\eta\right)=\frac{e^{-\eta}\eta^{N}}{N!},
\end{equation}
with $\sum_{N=0}^{\infty}P_{N}=1$ and $P_{0}=e^{-\eta}$. Accordingly,
the probability of absorbing at least one photon is $P_{N>0}\left(\eta\right)=1-e^{-\eta}$
and the probability of absorbing more than one photon per nanocrystals
is given by, $P_{N>1}\left(\eta\right)=1-e^{-\eta}-\eta e^{-\eta}$. 

Returning to an \emph{optically thick }sample where pump intensity
reduces significantly over the cell, all the above can be used to
calculate the density per unit area of particles which have absorbed
$N$ pump photons as:
\begin{equation}
\rho_{N}=\frac{1}{\sigma}\int_{\eta_{0}}^{\eta_{\infty}}P_{N}\left(\eta\right)\frac{d\eta}{\eta}.\label{eq:rho_N}
\end{equation}

Finally before deriving expressions for the fractional BE bleach we
require two more densities, $\rho_{N>0}$ the density per unit area
of dots absorbing at one or more photons, and $\rho_{N>1}$ the density
of those absorbing two or more photons: 

\begin{align}
\rho_{N>0} & =\frac{1}{\sigma}\int_{\eta_{0}}^{\eta_{\infty}}\left(1-e^{\eta}\right)\frac{d\eta}{\eta},\label{eq:rho_N>0}\\
\rho_{N>1} & =\frac{1}{\sigma}\int_{\eta_{0}}^{\eta_{\infty}}\left(1-e^{\eta}-\eta e^{\eta}\right)\frac{d\eta}{\eta}.\label{eq:rho_N>1}
\end{align}

To derive an expression for the intensity dependent fractional bleaching
at the band edge (BE) we use the equations for $\rho_{T}$, $\rho_{N>0}$
and $\rho_{N>1}$ along with $\eta_{\infty}$,$\eta_{0}$,and $\sigma_{BE}$
the cross section at the first exciton peak. Notice that when pumping
high in the inter-band continuum, $\sigma$ is often more than 10
times larger than $\sigma_{BE}$. Assuming linear bleaching of the
BE transition up to a degeneracy of 2, $\alpha_{BE}$ the sample band
edge absorbance after excitation (and carrier cooling) $\alpha_{BE,\eta_{0}}$
is obtained by integrating over the pump fluences throughout the cell:
\begin{equation}
\alpha_{BE,\eta_{0}}=\left(\rho_{T}-\frac{\rho_{N>0}}{2}-\frac{\rho_{N>1}}{2}\right)\sigma_{BE}.
\end{equation}
The fraction of residual absorption after pump excitation is accordingly,
\begin{align}
\frac{\alpha_{BE,\eta_{0}}}{\alpha_{BE}} & =\frac{2\rho_{T}-\rho_{N>0}-\rho_{N>1}}{2\rho_{T}}.\label{eq:alphaRatio}
\end{align}

Substituting Eqs.~(\ref{eq:rho_T}), (\ref{eq:rho_N})-(\ref{eq:rho_N>1})
into Eq.~(\ref{eq:alphaRatio}) we obtain:
\begin{equation}
\frac{\alpha_{BE,\eta_{0}}}{\alpha_{BE}}=\frac{1}{\log\left|\eta_{0}/\eta_{\infty}\right|}\int_{\eta_{\infty}}^{\eta_{0}}\left(e^{-\eta}+0.5\eta e^{-\eta}\right)\frac{d\eta}{\eta},
\end{equation}
from which the final fractional bleach is
\begin{equation}
\frac{\Delta\alpha_{BE,\eta_{0}}}{\alpha_{BE}}=1-\frac{1}{\log\left|\eta_{0}/\eta_{\infty}\right|}\int_{\eta_{\infty}}^{\eta_{0}}\left(e^{-\eta}+0.5\eta e^{-\eta}\right)\frac{d\eta}{\eta},
\end{equation}

and this is the expression used to simulate the plot of $\Delta\alpha/\alpha_{0}$vs
$\eta_{0}$in Fig.~5B of the manuscript.

\appendix

\section{\label{sec:The-Harmonic-correlation}Harmonic correlation function}

The time correlation function of the Harmonic Oscillator at temperature
$\beta$ is
\begin{align}
C\left(t,\beta\right) & =\frac{1}{m^{2}}\sum_{n}\frac{e^{-\beta E_{n}}}{Z}\left\langle n\left|\hat{p}\left(t\right)\hat{p}\left(0\right)\right|n\right\rangle \\
 & =\frac{1}{m^{2}}\sum_{nm}\frac{e^{-\beta E_{n}}}{Z}e^{iE_{nm}t}\left\langle n\left|\hat{p}\left|m\right\rangle \left\langle m\right|\hat{p}\right|n\right\rangle \\
 & =\frac{1}{m^{2}}\sum_{nm}\frac{e^{-\beta E_{n}}}{Z}e^{iE_{nm}t}\left|\left\langle n\left|\hat{p}\right|m\right\rangle \right|^{2}
\end{align}
where $M\hbar=P^{2}t$
\begin{equation}
Z\left(\beta\right)=\sum_{n=0}^{\infty}e^{-\beta\hbar\omega n}=\frac{1}{1-e^{-\beta\hbar\omega}}
\end{equation}
\begin{align}
\left\langle n\left|\hat{p}\right|m\right\rangle  & =i\sqrt{\frac{\hbar m\omega}{2}}\left\langle n\left|\hat{a}^{\dagger}-\hat{a}\right|m\right\rangle \\
 & =i\sqrt{\frac{\hbar m\omega}{2}}\left(\sqrt{m+1}\delta_{n,m+1}-\sqrt{m}\delta_{n,m-1}\right)
\end{align}
hence
\begin{equation}
\left|\left\langle n\left|\hat{p}\right|m\right\rangle \right|^{2}=\frac{m\hbar\omega}{2}\left(\left(m+1\right)\delta_{n,m+1}+m\delta_{n,m-1}\right)
\end{equation}
\begin{align}
C\left(t,\beta\right) & =\frac{\hbar\omega}{2m}\left(e^{i\omega t}+e^{-i\omega t}e^{\beta\hbar\omega}\right)\frac{1}{Z}\sum_{n=0}^{\infty}ne^{-\beta n\hbar\omega}\\
 & =\frac{\hbar\omega}{2m}\left(e^{i\omega t}+e^{-i\omega t}e^{\beta\hbar\omega}\right)\left\langle n\right\rangle _{\beta}
\end{align}
and
\begin{align}
\left\langle n\right\rangle _{\beta,\omega} & =\frac{1}{Z}\sum_{n=0}^{\infty}ne^{-\beta n\hbar\omega}\\
 & =-\frac{1}{\hbar\omega}\frac{\partial}{\partial\beta}\ln Z\\
 & =\frac{1}{\hbar\omega}\frac{\partial}{\partial\beta}\ln\left(1-e^{-\beta\hbar\omega}\right)\\
 & =\frac{1}{e^{\beta\hbar\omega}-1}
\end{align}
Clearly
\begin{align}
\left\langle n\right\rangle _{\beta,\omega}+1 & =e^{\beta\hbar\omega}\left\langle n\right\rangle _{\beta,\omega}
\end{align}
finally, 
\begin{align}
C\left(t,\beta\right) & =\frac{\hbar\omega}{2m}\left\langle n\right\rangle _{\beta\omega}\left(e^{i\omega t}+e^{-i\omega t}e^{\beta\hbar\omega}\right)\\
 & =\frac{\hbar\omega}{2m}\left(\left\langle n\right\rangle _{\beta\omega}e^{i\hbar\omega t}+e^{-i\hbar\omega t}\left(\left\langle n\right\rangle _{\beta\omega}+1\right)\right)
\end{align}
if we assume that the frequency is discrete $\omega_{m}=\frac{2\pi}{T_{d}}m$
where $m$ is integer then the response function of Eq.~(\ref{eq:bathResponse})
assumes the following form:
\begin{align}
\gamma\left(\pm\omega,\beta\right) & =T_{d}\frac{\hbar\omega}{2m}\left(\left\langle n\right\rangle _{\beta\omega}+\frac{1\pm1}{2}\right).\label{eq:gamma}
\end{align}

\onecolumngrid

\section{Cd36H36 Input file}

This is the Q-CHEM input file, including the nuclear configuration
and the basis set.

\texttt{}
\begin{lstlisting}
$molecule
0 1
Cd 2.26266696 0.80336407 -6.46510988
Cd 2.26266696 0.80336407 6.46510988
Cd -0.43389281 -2.49625428 -6.40155700
Cd -0.43389281 -2.49625428 6.40155700
Cd -1.88543621 1.45902174 -6.49895678
Cd -1.88543621 1.45902174 6.49895678
Cd -4.10780789 -1.20308722 -4.75283647
Cd -4.10780789 -1.20308722 4.75283647
Cd 0.83699863 4.07573258 -4.85280032
Cd 0.83699863 4.07573258 4.85280032
Cd 3.09910097 -2.83326887 -4.78577050
Cd 3.09910097 -2.83326887 4.78577050
Cd 2.83458308 0.82682770 -2.31844206
Cd 2.83458308 0.82682770 2.31844206
Cd -0.68852828 -3.06559991 -2.27167782
Cd -0.68852828 -3.06559991 2.27167782
Cd -2.12214994 1.94667024 -2.34781506
Cd -2.12214994 1.94667024 2.34781506
Cd -4.03854179 -1.39280936 0.00000000
Cd 3.40396851 -2.47907728 0.00000000
Cd 0.34766961 4.16957255 0.00000000
Cd 5.42916889 3.32370962 0.00000000
Cd -5.60011141 3.27040539 0.00000000
Cd 0.37762859 -6.53714690 0.00000000
Cd 4.57345019 3.90346038 -4.31177363
Cd 4.57345019 3.90346038 4.31177363
Cd -5.68257664 2.23505087 -4.30733764
Cd -5.68257664 2.23505087 4.30733764
Cd 1.23344351 -6.03770026 -4.29679356
Cd 1.23344351 -6.03770026 4.29679356
Cd 3.23078991 6.19612775 -2.10464336
Cd 3.23078991 6.19612775 2.10464336
Cd -7.18687857 0.06708793 -2.12106907
Cd -7.18687857 0.06708793 2.12106907
Cd 3.90521114 -5.97673874 -2.11084867
Cd 3.90521114 -5.97673874 2.11084867
Se -1.50427559 -4.48878986 0.00000000
Se 4.45358377 0.70162186 0.00000000
Se -2.80571365 3.38671429 0.00000000
Se 1.48190896 6.73882226 0.00000000
Se -6.81740331 -1.70124355 0.00000000
Se 5.16109123 -4.67106092 0.00000000
Se -2.63307265 -0.89017218 -2.35055978
Se -2.63307265 -0.89017218 2.35055978
Se 0.64055415 2.65897350 -2.37385332
Se 0.64055415 2.65897350 2.37385332
Se 1.95497191 -1.90482171 -2.34921571
Se 1.95497191 -1.90482171 2.34921571
Se 1.68463259 -7.73753003 -2.14176246
Se 1.68463259 -7.73753003 2.14176246
Se 5.83081241 5.07118510 -2.12602700
Se 5.83081241 5.07118510 2.12602700
Se -7.32389824 2.89605112 -2.15331364
Se -7.32389824 2.89605112 2.15331364
Se -1.08500310 -4.57016641 -4.66077668
Se -1.08500310 -4.57016641 4.66077668
Se 4.38648158 1.17716053 -4.69347086
Se 4.38648158 1.17716053 4.69347086
Se -3.14171439 3.20684327 -4.70192993
Se -3.14171439 3.20684327 4.70192993
Se 2.74601892 6.05192643 -4.87800129
Se 2.74601892 6.05192643 4.87800129
Se -6.73130484 -0.37782115 -4.85442287
Se -6.73130484 -0.37782115 4.85442287
Se 3.98914709 -5.42724163 -4.87277230
Se 3.98914709 -5.42724163 4.87277230
Se -2.72602690 -1.09340696 -7.16667357
Se -2.72602690 -1.09340696 7.16667357
Se 0.45328051 2.74250861 -7.26553396
Se 0.45328051 2.74250861 7.26553396
Se 2.14166044 -1.86454113 -7.21620233
Se 2.14166044 -1.86454113 7.21620233
$end
$rem
JOBTYPE sp
SCF_FINAL_PRINT 1
SCF_CONVERGENCE 6
PRINT_ORBITALS 1000
MOLDEN_FORMAT TRUE
basis gen
exchange PW91
SCF_ALGORITHM diis_gdm
CIS_N_ROOTS 100
CIS_SINGLETS true
CIS_TRIPLETS true
CIS_CONVERGENCE 6
CALC_SOC true
SYMMETRY false
UNRESTRICTED false
SYM_IGNORE true
max_scf_cycles 500
mem_total 200000
mem_static 2000
$end
$basis
Se 0 
S 3 1.00
2480.6268140 0.1543289673 
451.8492708 0.5353281423 
122.2880464 0.4446345422 
SP 3 1.00
206.1578780 -0.0999672292 0.1559162750 
47.90665727 0.3995128261 0.6076837186 
15.58073180 0.7001154689 0.3919573931 
SP 3 1.00
17.63999414 -0.2277635023 0.0049515112 
5.380760465 0.2175436044 0.5777664691 
2.076064666 0.9166769611 0.4846460366 
SP 3 1.00
1.2146442970 -0.3088441215 -0.1215468600 
0.4482801363 0.0196064117 0.5715227604 
0.1979652346 1.1310344420 0.5498949471 
D 3 1.00
17.63999414 0.2197679508 
5.380760465 0.6555473627 
2.076064666 0.2865732590 
****
Cd 0 
S 3 1.00
4950.2619050 0.1543289673 
901.6963856 0.5353281423 
244.0342313 0.4446345422 
SP 3 1.00
433.4469385 -0.0999672292 0.1559162750 
100.7237469 0.3995128261 0.6076837186 
32.75848861 0.7001154689 0.3919573931 
SP 3 1.00
52.59279235 -0.2277635023 0.0049515111 
16.04247800 0.2175436044 0.5777664691 
6.189686744 0.9166769611 0.4846460366 
SP 3 1.00
5.674851796 -0.3306100626 -0.1283927634 
2.209757875 0.0576109533 0.5852047641 
0.9727408566 1.1557874500 0.5439442040 
SP 3 1.00
0.5949150981 -0.3842642607 -0.3481691526 
0.3203250000 -0.1972567438 0.6290323690 
0.1414931855 1.3754955120 0.6662832743 
D 3 1.00
52.59279235 0.2197679508 
16.04247800 0.6555473627 
6.189686744 0.2865732590 
D 3 1.00
3.642963976 0.1250662138 
1.418551290 0.6686785577 
0.6244497700 0.3052468245 
****
$end
\end{lstlisting}

\twocolumngrid

\end{document}